\documentclass[onecolumn,preprint,showpacs,superscriptaddress]{revtex4-1}

\usepackage[utf8]{inputenc}
\usepackage{mathtools}
\usepackage{graphicx}
\usepackage{caption}
\usepackage{subcaption}
\usepackage{xcolor}
\usepackage{soul}
\usepackage{float}
\usepackage{amsmath}
\usepackage{amssymb}
\usepackage{amsfonts}

\begin{document}

\title{Scalar bosons in Bonnor-Melvin-$\Lambda$ universe: Exact solution, Landau levels and Coulomb-like potential}

\author{L. G. Barbosa}
\email{leonardo.barbosa@posgrad.ufsc.br}

\affiliation{Departamento de Física, CFM - Universidade Federal de \\ Santa Catarina; C.P. 476, CEP 88.040-900, Florianópolis, SC, Brazil}

\author{C. C. Barros Jr.}
\email{barros.celso@ufsc.br}

\affiliation{Departamento de Física, CFM - Universidade Federal de \\ Santa Catarina; C.P. 476, CEP 88.040-900, Florianópolis, SC, Brazil} 

\begin{abstract}
In this work, we study spin-0 particles in a spacetime whose structure is determined by a homogeneous magnetic field and a cosmological constant. For this purpose, we take into account a framework based on the Bonnor-Melvin solution with the inclusion of the cosmological constant. We write the Klein-Gordon equation, solve it, and determine the Landau levels. The effects of scalar and vector potentials are considered, and we investigate the influence of the parameters of the theory on the results, which present observable effects. The implications of the physics of a stellar model based on this framework are also discussed. 
\end{abstract}

\maketitle

\section{Introduction}
Probably one of the most intriguing theories in physics is the general theory of relativity. As far as it describes the gravitational interaction in terms of the geometry of the spacetime, many unexpected effects have been discovered such as the Lense-Thirring effect \cite{schiff1960possible}, \cite{lense1918einfluss}, that deals with the frame dragging, gravitational waves \cite{Abbott_2016} or even gravitational lensing \cite{einsteingl}, \cite{SRefsdal_1994}. These are just few examples among many others that may be found in the literature nowadays.

Recently, the attention in studies about magnetic fields has increased as many systems with very strong fields have been reported. In cosmological models, during the electroweak phase transition in the early universe, fields as strong as $10^{23}-10^{24}$ Gauss are supposed to be produced \cite{VACHASPATI1991258}. In non-central high energy heavy ion collisions, magnetic fields as high as $10^{18}$ G are expected to be produced at the Relativistic Heavy Ion Collider (RHIC) \cite{kharzeevmag}, \cite{BZDAK2012171}, \cite{voronyuk}, \cite{XU2020135706} and  $B\sim 10^{20}$ G   at the Large Hadron Collider (LHC) \cite{Skokov20095925}, \cite{Deng2012}. In the study of pulsars, from the measurement of the periods and period derivatives generating the $P-\dot{P}$ diagram, it was possible to estimate the magnetic fields of such stars and values of the order of $10^{10}-10^{12}$ G have been obtained \cite{Chatterjee_2021} at their surfaces. Further studies provided fields of the order of $B\sim 10^{16}-10^{16}$ G on the surface of the stars called  magnetars \cite{dunc1995}, 
\cite{Kouveliotou:1998ze}, \cite{Harding_2006} and estimates of a maximum magnetic field  $B\sim 10^{18}$ G in their interiors. 

The existence of a magnetic field, if strong enough, may significantly affect the structure of a star. The magnetic field interacts with charged particles inside the star (electrons, protons, hyperons, and others), determining the Landau levels and consequently modifying the equation of state. Strong magnetic fields also affect the energy-momentum tensor, breaking the spherical symmetry \cite{Chatterjee_2021}. It is also interesting to note that if the field exceeds the called quantum critical field strength $B=4.4\times 10^{13}$ G, which means that the cyclotron energy
$E=\hbar eB/m_ec$ exceeds the electron rest mass $m_e$, it affects significantly the physical processes and introduces different ones \cite{Harding_2006}. In \cite{chamel2015} it is shown that depending on the intensity of $B$, the neutron-drip density in the crust of the magnetar may be increased up to 14\% and decreased by 25\%. Even the emission of neutrons by the stars depends on the strength of this field \cite{Chamel_2016}.

So, a question of interest is how to incorporate the magnetic field in the general relativity framework.
Currently, some solutions of the Einstein-Maxwell equations are known, as for example the Manko solution \cite{GUTSUNAEV1987215}, \cite{GUTSUNAEV198885}, the Bonnor-Melvin universe \cite{WBBonnor_1954}, \cite{MELVIN196465} and a formulation based on the Bonnor-Melvin solution with the inclusion of the cosmological constant \cite{vzofka2019bonnor}. In this work, the first objective is to derive a variation of the solution presented in \cite{vzofka2019bonnor}, making explicit the role of the cosmological constant $\Lambda$ and the process of limiting the solution in a region near astrophysical objects.

Still thinking about the general relativity, another important aspect is how this theory may be related with the quantum physics, or if this kind of relation may also exist, or at least be relevant. At the moment many works have been proposed with this purpose and are essentially based on the Klein-Gordon and Dirac equations written in curved spacetimes \cite{Parker:1980hlc}. Typical examples are particles
in the Schwarzschild \cite{elizalde_1987}, Kerr black holes \cite{chandra} and in cosmic string \cite{Santos:2016omw}, \cite{Santos:2017eef}, \cite{Vitoria:2018its} backgrounds, quantum oscillators \cite{Ahmed:2022tca}, \cite{Ahmed:2023blw}, \cite{Santos:2019izx}, \cite{Yang:2021zxo}, \cite{Soares:2021uep}, \cite{Rouabhia:2023tcl},
 Casimir effect \cite{Bezerra:2016brx}, \cite{Santos:2018jba} or particles in the Hartle-Thorne spacetime \cite{Pinho:2023nfw}. These studies and others \cite{Cano:2021qzp}, \cite{Sedaghatnia:2019xqb}, \cite{Guvendi:2022uvz},
\cite{Vitoria:2018mun}
have provided many interesting results and insights about how quantum systems are affected by arbitrary geometries of the spacetime by calculating the wave functions, energy levels and other quantum effects.

So, an interesting subject is to study quantum particles in a spacetime with the structure affected by a magnetic field in a way similar to the one done in \cite{Santos:2015esa}, where Dirac particles have been studied in the Melvin metric. In this paper, we will study spin-0 bosons in the magnetic universe with cosmological constant, proposing a variation in the constants of the metric shown in \cite{vzofka2019bonnor} and also verify the effect of the inclusion of a vector potential $A_{t}$ and a scalar potential $S(r)=\eta/r$. As it was pointed out before, the Landau levels play a fundamental role in the determination of the equation of state of the magnetars, and for this reason, they will be calculated in the backgrounds determined by this kind of interaction and the effect of the parameters of the theory will be investigated.

The structure of this paper is as follows: In Sec. II we review the Bonnor-Melvin-$\Lambda$ and propose a different definition of the constants. In Sec. III, the Klein-Gordon equation is obtained and solved in this spacetime, and then the Landau levels are determined. In Sec. IV a scalar Coulomb-like potential is introduced, leading to a new equation that is solved, and in Sec. V we draw our conclusions.

\section{Bonnor-Melvin-$\Lambda$ universe}

In this section we will review the solution presented in \cite{vzofka2019bonnor} for the Bonnor-Melvin with cosmological constant ($\Lambda$) spacetime and propose a variation of the definition of the integration constant in order to have an explicitly well-behaved solution in the limit $\Lambda\rightarrow 0$.
To begin, let's examine Einstein's field equations in the presence of a positive cosmological constant 
\begin{equation}\label{EqCamp}
R_{\mu\nu}-\frac{1}{2}Rg_{\mu\nu}-\Lambda g_{\mu\nu}=-8\pi T_{\mu\nu}
\end{equation}
where we consider the electromagnetic energy-momentum tensor
\begin{equation}
T_{\mu\nu}=\frac{1}{4\pi}\left(F_{\mu}^{\quad\alpha}F_{\nu\alpha}-\frac{1}{4}F_{\alpha\beta}F^{\alpha\beta}g_{\mu\nu}\right),
\end{equation}
in which $F_{\mu\nu}=\nabla_{\nu}A_{\mu}-\nabla_{\mu}A_{\mu}$ and obeys the Maxwell equation $\nabla_{\mu}F^{\mu\nu}=0$. We are interested in obtaining a static solution with cylindrical symmetry, to do so we suppose that the line element has the form
\begin{equation}
    ds^{2}=-e^{A\left(r\right)}dt^{2}+dr^{2}+e^{B\left(r\right)}dz^{2}+e^{C\left(r\right)}d\varphi^{2}
\end{equation}
where $t,z\in\mathbb{R}$, $r\in\mathbb{R}_{+}$, and $\varphi\in\left[0,2\pi\right)$, and as it has been shown in \cite{vzofka2019bonnor}, 
$A\left(r\right)$ and $B\left(r\right)$ are constants, and then this line element may be written as
\begin{equation}
  ds^{2}=-dt^{2}+dr^{2}+dz^{2}+e^{C\left(r\right)}d\varphi^{2}.
\end{equation}

By assuming a purely magnetic field aligned with the axis of symmetry, generated by an electromagnetic potential of the form $A_{\mu}=\left(0,0,0,A_{\varphi}\left(r\right)\right)$ such that $F_{r\varphi}=\partial_{r}A_{\varphi}=H\left(r\right)$, from the field equations (\ref{EqCamp}) we obtain
\begin{equation}\label{EqCamp1}
   \frac{d^{2}C\left(r\right)}{d r^{2}}+\frac{1}{2}\left(\frac{d C\left(r\right)}{d r}\right)^{2}+\left(2\Lambda+2e^{-C(r)}H^{2}\left(r\right)\right)=0
\end{equation}
\begin{equation}\label{EqCamp2}
e^{-C(r)}H^{2}\left(r\right)-\Lambda=0,
\end{equation}
substituting (\ref{EqCamp2}) into (\ref{EqCamp1}), we obtain
\begin{equation}\label{EqCamp3}
\frac{d^{2}C\left(r\right)}{d r^{2}}+\frac{1}{2}\left(\frac{d C\left(r\right)}{dr}\right)^{2}+4\Lambda=0
\end{equation}
which is the same equation found in \cite{vzofka2019bonnor}. A physically reasonable solution to equation (\ref{EqCamp3}) is given by
\begin{equation}\label{newsolution}
C\left(r\right)=\ln\left(\frac{\sigma^{2}}{2\Lambda}\sin^{2}\left(\sqrt{2\Lambda}r\right)\right) ,
\end{equation}
where $\sigma$ is a dimensionless integration constant that we take as a positive real number, and leads to the following line element
\begin{equation}\label{metrica}
ds^{2}=-dt^{2}+dr^{2}+dz^{2}+\frac{\sigma^{2}}{2\Lambda}\sin^{2}\left(\sqrt{2\Lambda}r\right)d\varphi^{2}.
\end{equation}
Mathematically eq. (\ref{newsolution}) may be viewed as the same solution presented in \cite{vzofka2019bonnor}, but in this work we defined the integration constant in a different form, presenting an explicit dependence on the cosmological constant in a way to provide an easier interpretation of the results. 

In order to determine the electromagnetic potential and, consequently, the magnetic field, we use a Maxwell's equation, which can be written as
\begin{equation}
\frac{1}{\sqrt{-g}}\partial_{\nu}\left(\sqrt{-g}g^{\alpha\mu}g^{\beta\nu}F_{\alpha\beta}\right)=0 .
\end{equation}
Solving the equation, we obtain an expression for the electromagnetic potential
\begin{equation}
A_{\varphi}=-\sigma\cos\left(\sqrt{2\Lambda}r\right) ,
\end{equation}
and then we can write the magnetic field as
\begin{equation}
H\left(r\right)=\sigma\sqrt{2\Lambda}\sin\left(\sqrt{2\Lambda}r\right).
\end{equation}

We should note that the line element and the magnetic field have appropriate units in the geometric unit system ($G=c=1$) and both are well-defined for $\Lambda\rightarrow 0$, if we consider regions near typical astrophysical objects, such as stars and even galaxies, which means small values of $r$ if compared with cosmological scales, we have    
\begin{equation}
\lim_{\Lambda\rightarrow0}g_{\varphi\varphi}=\lim_{\Lambda\rightarrow0}\frac{\sigma^{2}}{2\Lambda}\sin^{2}\left(\sqrt{2\Lambda}r\right)=\sigma^{2}r^{2},
\end{equation}
that plays the role of a deficit angle, similar to the one that appears when studying cosmic strings \cite{Santos:2016omw}, \cite{Santos:2017eef}.

Defining the magnetic field intensity as $H_{0}=\sigma\sqrt{2\Lambda}$, the integration constant can be characterized as $\sigma=H_{0}/\sqrt{2\Lambda}$, so that we can rewrite the line element and the magnetic field in terms only of the cosmological constant and the magnetic field
\begin{equation}
ds^{2}=-dt^{2}+dr^{2}+dz^{2}+\frac{H_{0}^{2}}{4\Lambda^{2}}\sin^{2}\left(\sqrt{2\Lambda}r\right)d\varphi^{2} \ ,
\label{dsh}
\end{equation}
\begin{equation}
H\left(r\right)=H_{0}\sin\left(\sqrt{2\Lambda}r\right)
\  ,
\end{equation}
which characterizes a solution of the Einstein-Maxwell equations with a positive cosmological constant that generates and maintains a homogeneous magnetic field. This is a spacetime of constant curvature, and in the limit of a vanishing cosmological constant with $\sigma=1$, that also means $H_0\rightarrow 0$ and the Ricci scalar $R\rightarrow 0$, it is flat spacetime written in cylindrical coordinates. The reason to define the constants as it was done in this section was to obtain this kind of behavior explicitly in the metric.

\section{Klein-Gordon equation in the Bonnor-Melvin-$\Lambda$ universe} 
In this work we are interested in studying charged scalar bosons in the Bonnor-Melvin spacetime with cosmological constant, and for this purpose we must notice that these particles can be described by the Klein-Gordon equation in a covariant form \cite{Parker:1980hlc} that is given by (with $\hslash=c=1$)
\begin{equation} \label{eq:2}
  \frac{1}{\sqrt{-g}}\left(\partial_{\mu}-ieA_{\mu}\right)\left(\sqrt{-g}g^{\mu\nu}\left(\partial_{\nu}-ieA_{\nu}\right)\Psi\right)-m^{2}\Psi=0,
\end{equation}
where $e$ is the charge of the particle and $m$ is its mass. In terms of the line element (\ref{metrica}) the equation is
\begin{equation}
 g^{tt}\frac{\partial^{2}\Psi}{\partial t^{2}}+\frac{1}{\sqrt{-g}}\frac{\partial}{\partial r}\left(g^{rr}\sqrt{-g}\frac{\partial\Psi}{\partial r}\right)+g^{zz}\frac{\partial^{2}\Psi}{\partial z^{2}}+g^{\varphi\varphi}\left(\frac{\partial}{\partial\varphi}-ieA_{\varphi}\right)^{2}\Psi-m^{2}\Psi=0
\end{equation}
and may be written as
\begin{equation} \label{eq:3}
\begin{split}
 &-\frac{\partial^{2}\Psi}{\partial t^{2}}+\frac{1}{\sin\left(\sqrt{2\Lambda}r\right)}\frac{\partial}{\partial r}\left(\sin\left(\sqrt{2\Lambda}r\right)\frac{\partial\Psi}{\partial r}\right)\\&+\frac{\partial^{2}\Psi}{\partial z^{2}}+\frac{2\Lambda}{\sigma^{2}\sin^{2}\left(\sqrt{2\Lambda}r\right)}\left(\frac{\partial}{\partial\varphi}-ieA_{\varphi}\right)^{2}\Psi-m^{2}\Psi=0.
\end{split}
\end{equation}
We should note that equation (\ref{eq:3}) does not depend on the variables $t$ and $z$, then the solution have the form
\begin{equation} \label{eq:4}
  \Psi\left(t,r,\varphi,z\right)=e^{-i\varepsilon t}e^{i\ell\varphi}e^{ip_{z}z}R\left(r\right)
\end{equation}
where $\varepsilon$, $\ell=0,\pm1,\pm2,\pm3,\cdots$ and $p_z$ are the quantum numbers associated with energy, angular momentum around $\varphi$ and momentum in the $z$ direction respectively,
and it remains for us to determine the dynamics of the radial component of the equation. Substituting (\ref{eq:4}) into (\ref{eq:3}) and defining
\begin{equation} \label{eq:5}
\xi^{2}=\varepsilon^{2}-m^{2}-p_{z}^{2}\end{equation}
 we get the radial differential equation 
\begin{equation} \label{eq:6}
 \frac{1}{\sin\left(\sqrt{2\Lambda}r\right)}\frac{d}{dr}\left[\sin\left(\sqrt{2\Lambda}r\right)\frac{dR\left(r\right)}{dr}\right]+\left[\xi^{2}-\frac{2\Lambda}{\sigma^{2}\sin^{2}\left(\sqrt{2\Lambda}r\right)}\left(\ell-eA_{\varphi}\right)^{2}\right]R\left(r\right)=0
\end{equation}
and to solve it we make the change of variables $u=\cos\left(\sqrt{2\Lambda}r\right)$, which results in
\begin{equation}
\left(1-u^{2}\right)\frac{d^{2}R\left(u\right)}{du^{2}}-2u\frac{dR\left(u\right)}{du}+\left[\frac{\xi^{2}}{2\Lambda}-\frac{1}{\left(1-u^{2}\right)}\left(\frac{\ell}{\sigma}+eu\right)^{2}\right]R\left(u\right)=0.
\end{equation}

Taking the following proposed solution for the radial function
\begin{equation}
R\left(u\right)=\left(u+1\right)^{-\frac{1}{2}e}\left(u-1\right)^{\frac{1}{2}e}\left(u^{2}-1\right)^{\frac{\left|\ell\right|}{2\sigma}}F\left(u\right)
\end{equation}
and then performing the variable change $z=\frac{1}{2}\left(u+1\right)$, we can express the radial equation in the form of a hypergeometric differential equation
\begin{equation}
z\left(1-z\right)\frac{d^{2}F\left(z\right)}{dz^{2}}+\left[c-\left(a+b+1\right)z\right]\frac{dF\left(z\right)}{dz}-abF\left(z\right)=0
\end{equation}
whose parameters $a$, $b$ and $c$ given respectively by $a=\left|\ell\right|/\sigma+\left(2\xi^{2}/\Lambda+4e^{2}+ 1\right)^{\frac{1}{2}}/2+1/2$, $b=\left|\ell\right|/\sigma - \left(2\xi^{2}/\Lambda +4e^{2}+1\right)^{\frac{1}{2}}/2+1/2$ and $c=1-e+\left|\ell\right|/\sigma$. The general solution of the hypergeometric equation can be given by
\begin{equation}
    F\left(z\right)=C_{1}{}_{2}F_{1}\left(a,b;c;z\right)+C_{2}z^{1-c}{}_{2}F_{1}\left(a+1-c,b+1-c;2-c;z\right)
\end{equation}
where $C_{1}$ and $C_{2}$ are constants of integration. Since we are interested in a well-defined solution at the origin, then we take $C_2=0$ and $ _{2}F_{1}\left(a,b;c;z\right)$ is the regular hypergeometric function,
\begin{equation}
  _{2}F_{1}\left(a,b;c;z\right)=\sum_{n=0}^{\infty}A_{n}z^{n}=\sum_{n=0}^{\infty}\frac{\left(a\right)_{n}\left(b\right)_{n}}{\left(c\right)_{n}}\frac{z^{n}}{n!}
\end{equation}
where the coefficients $A_{n}$ obey the recurrence relation 
\begin{equation}
    A_{n+1}=\frac{\left(n+a\right)\left(n+b\right)}{\left(n+1\right)\left(n+c\right)}A_{n}.
\end{equation}
From this recurrence relation, we have that the function $_{2}F_{1}\left(a,b;c;z\right)$ becomes a polynomial of degree $n$ for the following condition $a=-n$, $n=0,1,2,\cdots.$

Seeking to determine the energy spectrum of the charged scalar boson, we can consider the condition of a polynomial solution for the hypergeometric function, allowing us to obtain
\begin{equation}
\varepsilon_{\pm}=\pm\sqrt{m^{2}+p_{z}^{2}+2\Lambda\left[n\left(n+\frac{2\left|\ell\right|}{\sigma}+1\right)+\frac{\left|\ell\right|}{\sigma}\left(\frac{\left|\ell\right|}{\sigma}+1\right)-e^{2}\right]}
\end{equation}
and with the aim of expressing this energy spectrum in terms of physical quantities, we recall the relationship between the magnetic field intensity and the cosmological constant, $\sigma=H_{0}/\sqrt{2\Lambda}$, which implies
\begin{equation}
\varepsilon_{\pm}=\pm\sqrt{m^{2}+p_{z}^{2}+2\Lambda\left[n\left(n+\frac{2\sqrt{2\Lambda}\left|\ell\right|}{H_{0}}+1\right)+\frac{\sqrt{2\Lambda}\left|\ell\right|}{H_{0}}\left(\frac{\sqrt{2\Lambda}\left|\ell\right|}{H_{0}}+1\right)-e^{2}\right]} ,
\end{equation}

We must remark that up to this point, we have considered $G=c=\hslash=1$ and with the objective of expressing the energy in GeV, we make the conversion to the natural system of units that gives
\begin{equation}
\varepsilon_{\pm}=\pm\sqrt{m^{2}+p_{z}^{2}+2\Lambda\left[n\left(n+\frac{2\sqrt{2\Lambda}\left|\ell\right|}{\sqrt{G}H_{0}}+1\right)+\frac{\sqrt{2\Lambda}\left|\ell\right|}{\sqrt{G}H_{0}}\left(\frac{\sqrt{2\Lambda}\left|\ell\right|}{\sqrt{G}H_{0}}+1\right)-e^{2}\right]}
\end{equation}

\begin{figure}[H]
     \centering
     \begin{subfigure}[b]{0.3\textwidth}
         \centering
         \includegraphics[width=\textwidth]{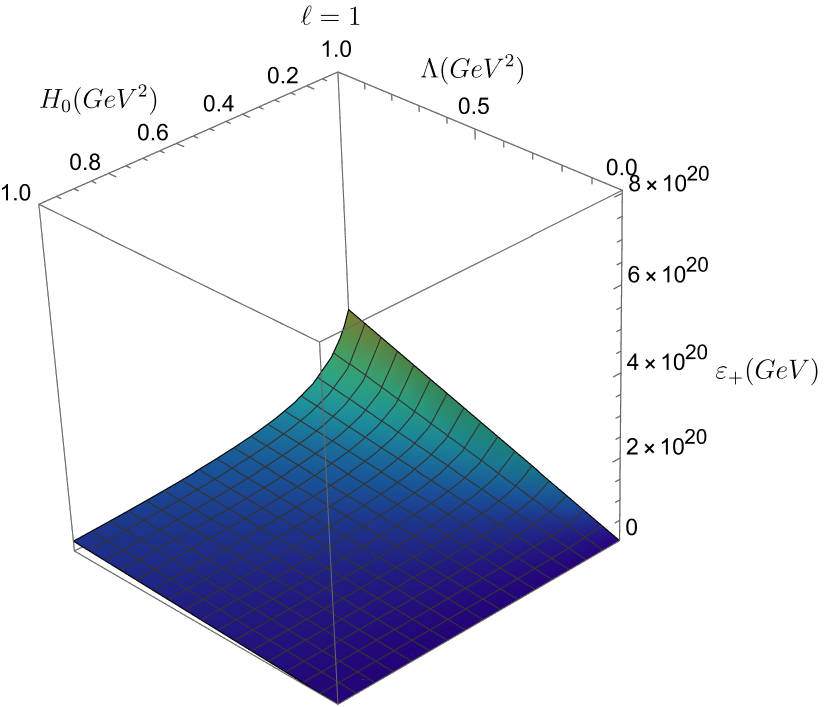}
     \end{subfigure}
     \begin{subfigure}[b]{0.3\textwidth}
         \centering
         \includegraphics[width=\textwidth]{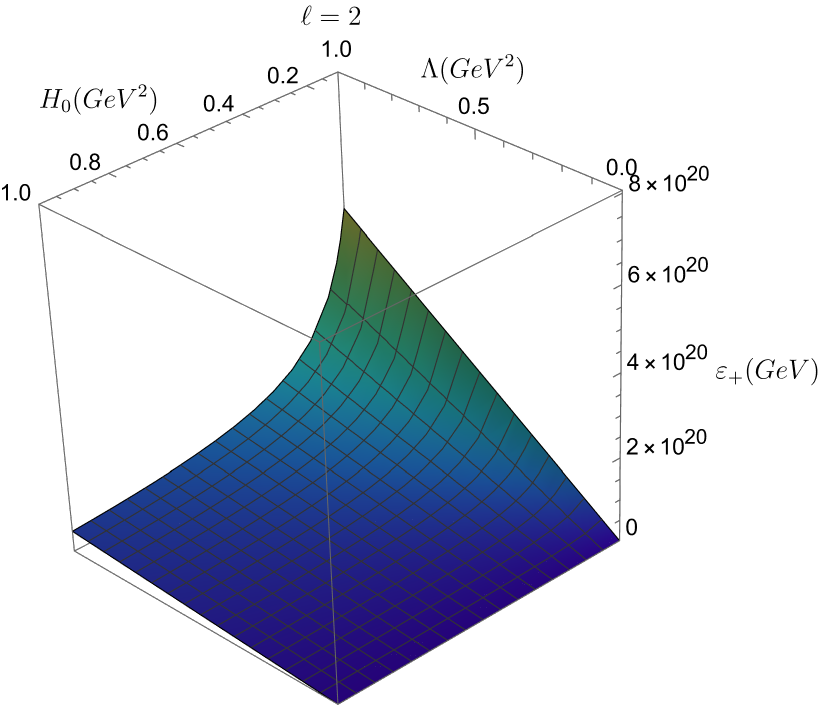}
     \end{subfigure}
    \begin{subfigure}[b]{0.3\textwidth}
         \centering
         \includegraphics[width=\textwidth]{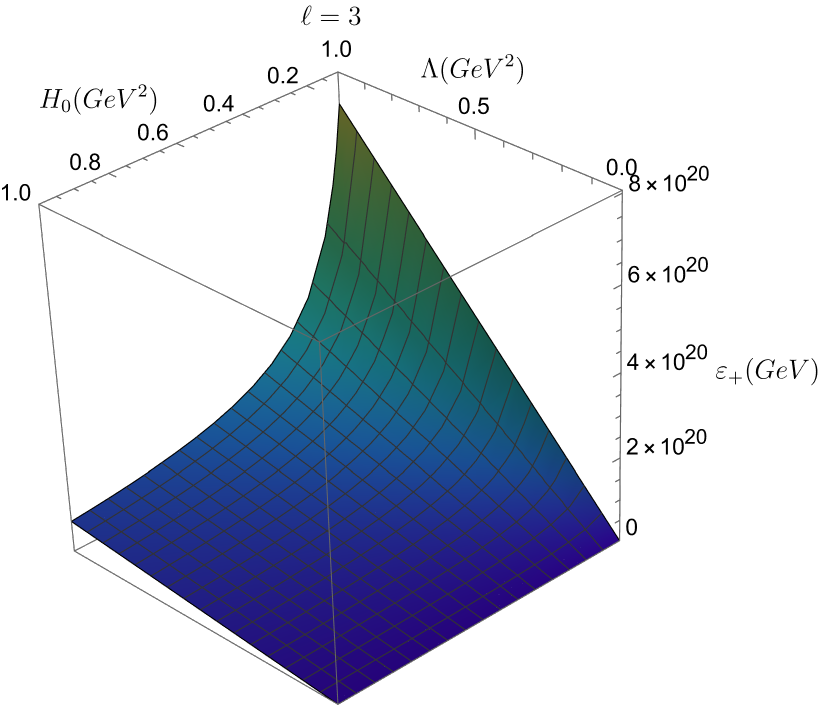}
     \end{subfigure}
        \caption{Graph of the energy spectrum in GeV with respect to the magnetic field intensity $H_{0}$ and the cosmological constant $\Lambda$, for $\ell=1,2,3$, $m=p_z= 0.139$ GeV, $G=6.70 \times 10^{-39}\text{GeV}^{2}$, $e=0.3$ and $n=1000$.}
        \label{f1}
\end{figure}

\begin{figure}[H]
     \centering
     \begin{subfigure}[b]{0.3\textwidth}
         \centering
         \includegraphics[width=\textwidth]{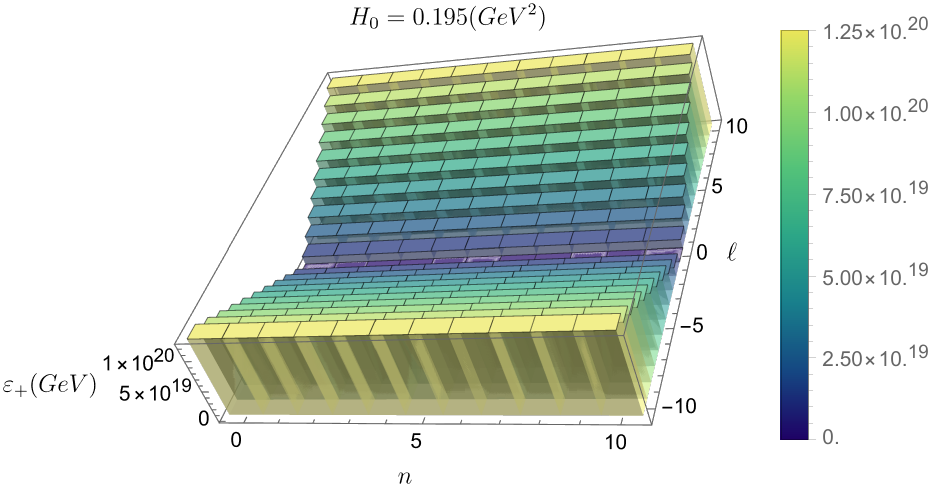}
     \end{subfigure}
     \begin{subfigure}[b]{0.3\textwidth}
         \centering
         \includegraphics[width=\textwidth]{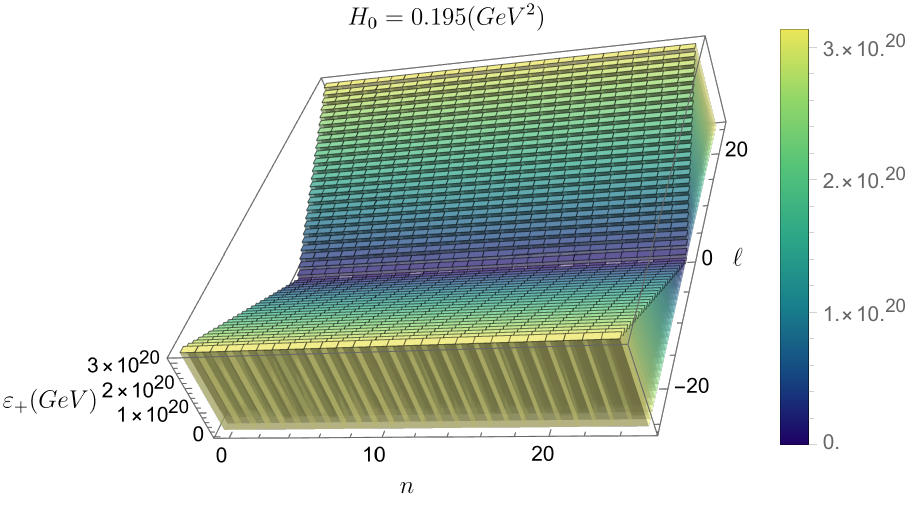}
     \end{subfigure}
    \begin{subfigure}[b]{0.3\textwidth}
         \centering
         \includegraphics[width=\textwidth]{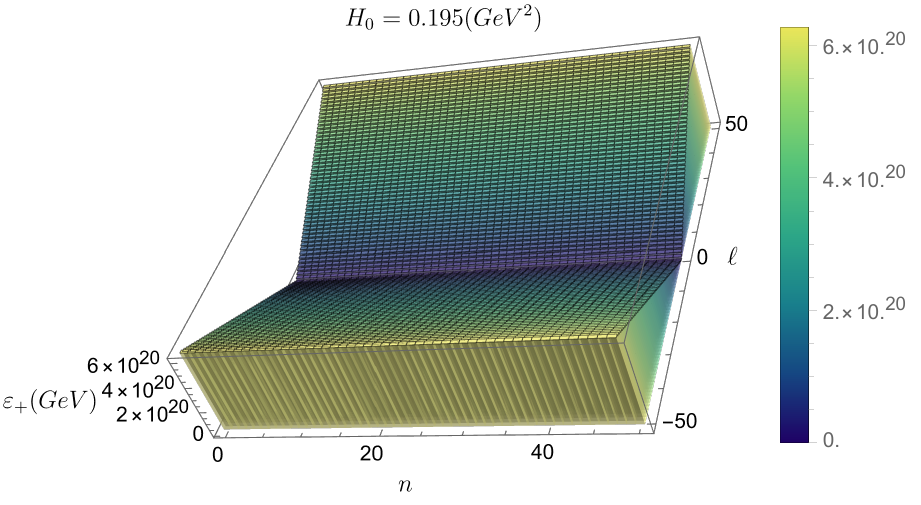}
     \end{subfigure}
        \caption{Plot of the energy spectrum in GeV with respect to the quantum numbers $n$ and $\ell$, with $H_{0}=0.195 \text{ GeV}^{2}$ and the cosmological constant $\Lambda=0.1 \text{ GeV}^{2}$, for $m=p_z= 0.139$ GeV, $e=0.3$.}
        \label{f2}
\end{figure}

Figures \ref{f1} and \ref{f2} show the behavior of the energy spectrum considering a $\pi^{+}$ as a test particle with mass $m$=0.139 GeV \cite{Workman:2022ynf}, \cite{ kharzeevmag}. In Fig. \ref{f1} we have the variation of $\Lambda$ and $H_0$ for $n$=1000 as an example. As we can see, as $\Lambda\rightarrow 0$ for fixed values of $n$ the energy goes to $\varepsilon\sim\sqrt{m^2+p_z^2}$ and describes free particles. The value of the energy levels increase as $\Lambda$ increases and also increases for smaller values of $H_0$. This fact may be understood observing the behavior of $\sigma$, and the structure of the spacetime, that becomes similar to the cosmic string one, presenting a deficit angle \cite{Santos:2016omw}, \cite{Santos:2017eef}.

In Fig. \ref{f2}, the energy levels are shown in terms of the variation of the quantum numbers $n$ and $\ell$ for fixed values of $H_0$ and $\Lambda$, and as we can see this variation is important in the range considered in 
 the figure. 

The Klein-Gordon equation for charged scalar bosons in the Bonnor-Melvin-$\Lambda$ universe was solved, leading to a quantized energy spectrum. Figure \ref{f1} shows the energy levels as a function of the cosmological constant $\Lambda$ and the magnetic field intensity $H_0$ for a fixed quantum number $n = 1000$. When $\Lambda$ approaches zero, the energy levels approach those of free particles, indicating that the cosmological constant modifies the spectrum. As $\Lambda$ increases, the energy levels rise, showing the influence of the cosmological constant. For smaller values of $H_0$, the energy levels also increase, reflecting the role of the magnetic field in the quantization process. Figure \ref{f2} displays the energy levels as a function of $n$ and $\ell$ for fixed $\Lambda$ and $H_0$. The energy increases with both $n$ and $\ell$, demonstrating the dependence of the spectrum on the quantum numbers. These results highlight the quantization of energy levels (Landau levels) due to the magnetic field and the impact of $\Lambda$ on the spectrum.

\section{Coulomb-like potential}
Another system of interest is the one formulated in the same metric with the inclusion of a scalar potential. If the potential goes to zero for large values of $r$, it is reasonable to look for solutions for small values of $r$, that represent particles near the symmetry axis. Quantum particles bounded in small regions near the origin is a formulation that is useful in many physical problems, so they will be studied in this section. In this case we may expand eq. (\ref{dsh}) as
\begin{equation}
ds^{2}=-dt^{2}+dr^{2}+dz^{2}+\frac{H_{0}^{2}}{4\Lambda^{2}}\left[\sqrt{2\Lambda}r+\frac{\left( \sqrt{2\Lambda}r \right)^{3}}{3}+...\right]^{2}d\varphi^{2}
\label{dsex}
\end{equation}
that is
\begin{equation}
ds^{2}=-dt^{2}+dr^{2}+dz^{2}+\frac{H_{0}^{2}}{2\Lambda}r^{2}d\varphi^{2}
\end{equation}
for small values of $r$. If one considers the usual value $\Lambda=1.1 \times 10^{-52}\text{m}^{-2}$ and $r=1$ km, that is a large value for quantum systems, the first correction term in the metric given in eq. (\ref{dsex}) is $10^{-20}$ times smaller, so, to neglect this term and the subsequent ones in the expansion is a very good approximation.

A scalar potential $S\left(r\right)$ may be taken into account by making a modification on the mass term \cite{dosch1971kleins}, \cite{bergerhoff1994scalar}, \cite{figueiredo2012relativistic} $m\rightarrow m+S\left(r\right)$. Substituting this mass term in Klein-Gordon equation we obtain the following differential equation
\begin{equation}
    \frac{1}{\sqrt{-g}}\left(\partial_{\mu}-ieA_{\mu}\right)\left(g^{\mu\nu}\sqrt{-g}\left(\partial_{\nu}-ieA_{\nu}\right)\Psi\right)-\left(m+S\right)^{2}\Psi=0,
\end{equation}
then, as done previously, the solution for the wave equation may be written as $\Psi\left(t,r,z,\varphi\right)=e^{-i\varepsilon t}e^{i\ell\varphi}e^{ip_{z}z}R\left(r\right)$, which results in the radial equation
\begin{equation}
  \begin{split}
      &\frac{1}{\sin\left(\sqrt{2\Lambda}r\right)}\frac{d}{dr}\left[\sin\left(\sqrt{2\Lambda}r\right)\frac{dR\left(r\right)}{dr}\right]\\&+\left[\varepsilon^{2}-p_{z}^{2}-\frac{2\Lambda}{\sigma^{2}\sin^{2}\left(\sqrt{2\Lambda}r\right)}\left(\ell-eA_{\varphi}\right)^{2}-\left(m+S\right)^{2}\right]R\left(r\right)=0
  \end{split}
\end{equation}
which is still an equation obtained without any approximation. As an example we choose a Coulomb-type potential $S=\eta/r$, where $\eta$ is a coupling constant and then assuming a small value for the cosmological constant and solutions for small values of $r$, we can consider an expansion up to $\mathcal{O}\left( 2\Lambda r^{2}\right)$ as it was discussed before, so that the radial equation takes the form
\begin{equation}\label{EqRadialII}
    \frac{d^{2}R\left(r\right)}{dr^{2}}+\frac{1}{r}\frac{dR\left(r\right)}{dr}+\left[\varepsilon^{2}-p_{z}^{2}-\frac{1}{\sigma^{2}r^{2}}\left(\ell+e\sigma\left(1-\Lambda r^{2}\right)\right)^{2}-\left(m+\frac{\eta}{r}\right)^{2}\right]R\left(r\right)=0. 
\end{equation}
By assuming $R(r) = r^{-\frac{1}{2}} \Phi(r)$, we can rewrite equation (\ref{EqRadialII}) in the form of an effective Schrödinger equation with an effective potential $V_{\text{eff}}$
\begin{equation}
    \frac{d^{2}\Phi\left(r\right)}{dr^{2}}+\left(\xi^{2}-V_{\text{eff}}\right)\Phi\left(r\right)=0,
\end{equation}
\begin{equation}
  V_{\text{eff}}=-2\Lambda e\left(\frac{\ell}{\sigma}+e\right)+\frac{2m\eta}{r}+\left[\left(\frac{\ell}{\sigma}+e\right)^{2}+\eta^{2}-\frac{1}{4}\right]\frac{1}{r^{2}}+e^{2}\Lambda^{2}r^{2}.
\end{equation}

On the other hand, looking for an explicit solution for (\ref{EqRadialII}), we take the following solution proposal
\begin{equation}
    R\left(r\right)=r^{\sqrt{\left(\frac{\ell}{\sigma}+e\right)^{2}+\eta^{2}}}e^{-\frac{e\Lambda}{2}r^{2}}H\left(r\right)
\end{equation}
and then making the change of variable $x=\sqrt{e\Lambda}r$ the equation takes the form of a biconfluent Heun equation
\begin{equation}
    \frac{d^{2}H\left(x\right)}{dx^{2}}+\left(\frac{1+\alpha}{x}-\beta-2x\right)\frac{dH\left(x\right)}{dx}+\left\{ \left(\gamma-\alpha-2\right)-\frac{1}{2}\left[\delta+\left(1+\alpha\right)\beta\right]\frac{1}{x}\right\} H\left(x\right)=0
\end{equation}
whose parameters are $\alpha=2\left( \left(\ell/\sigma+e\right)^{2}+\eta^{2}\right)^{\frac{1}{2}}$, $\beta=0$, $\gamma =\left(\xi^{2}+2\Lambda e\left(\frac{\ell}{\sigma}+e\right)\right)/e\Lambda$ and $\delta=4m\eta/ \sqrt{e\Lambda}$. The general solution for this equation is given in terms of the bifluent Heun function \cite{ronveaux1995heun,Cunha:2016uch} 
\begin{equation}
    H\left(x\right)=C_{1}\text{HeunB}\left(\alpha,\beta,\gamma,\delta,x\right)+C_{2}x^{-\alpha}\text{HeunB}\left(-\alpha,\beta,\gamma,\delta;x\right)
\end{equation}
where $C_{1}$ and $C_{2}$ are constants, and as we are interested in a well-defined solution at the origin, we take $C_2=0$. For the case where $\alpha$ is not a negative integer, the biconfluent Heun functions can be written as
\begin{equation}
    \text{HeunB}\left(\alpha,\beta,\gamma,\delta;x\right)=\sum_{j=0}^{\infty}\frac{A_{j}}{\left(1+\alpha\right)_{j}}\frac{x^{j}}{j!}=\sum_{j=0}^{\infty}\frac{\Gamma\left(1+\alpha\right)A_{j}}{\Gamma\left(j+1+\alpha\right)}\frac{x^{j}}{j!}
\end{equation}
where the coefficients $A_{j}$ obey the recurrence relation $\left(j\geq0\right)$\begin{equation}
    A_{j+2}=\left[\left(j+1\right)\beta+\frac{1}{2}\left[\delta+\left(1+\alpha\right)\beta\right]\right]A_{j+1}-\left(j+1\right)\left(j+1+\alpha\right)\left(\gamma-\alpha-2-2j\right)A_{j}. 
\end{equation}
Observing this recurrence relation we can conclude that the $\text{HeunB}\left(\alpha,\beta,\gamma,\delta;x\right)$ function becomes a polynomial of degree $n$ if the following conditions are satisfied
\begin{equation}\label{HeunCondI}
    \gamma-\alpha-2=2n,\quad n=0,1,2,\cdots 
\end{equation}
\begin{equation}
    A_{n+1}=0
\end{equation}
where $A_{n+1}$ has $n+1$ real roots when 1$+\alpha>0$ and $\beta\in\mathbb{R}$. This condition may be represented by a three-dimensional diagonal determinant $\left(n+1\right)$ as it is shown in 
 \cite{ronveaux1995heun,vieira2015quantum}.

Condition (\ref{HeunCondI}) allows us to determine the energy spectrum through a quantization condition, which allows us to write
\begin{equation}
    \varepsilon_{\pm}=\pm\sqrt{m^{2}+p_{z}^{2}+2\Lambda e\left[n+\left(\left(\frac{\ell}{\sigma}+e\right)^{2}+\eta^{2}\right)^{\frac{1}{2}}-\left(\frac{\ell}{\sigma}+e\right)+1\right]} \ .
\end{equation}

As it was done previously, we intend to write the energy spectrum in terms of physical quantities, we then take $\sigma=H_{0}/\sqrt{2\Lambda}$, and recovering the gravitational constant $G$, we can write

\begin{equation}
   \varepsilon_{\pm}=\pm\sqrt{m^{2}+p_{z}^{2}+2\Lambda e\left[n+\left(\left(\sqrt{2\Lambda}\frac{\ell}{\sqrt{G}H_{0}}+e\right)^{2}+\eta^{2}\right)^{\frac{1}{2}}-\left(\sqrt{2\Lambda}\frac{\ell}{\sqrt{G}H_{0}}+e\right)+1\right]} \  .
\end{equation}

\begin{figure}[H]
     \centering
     \begin{subfigure}[b]{0.3\textwidth}
         \centering
         \includegraphics[width=\textwidth]{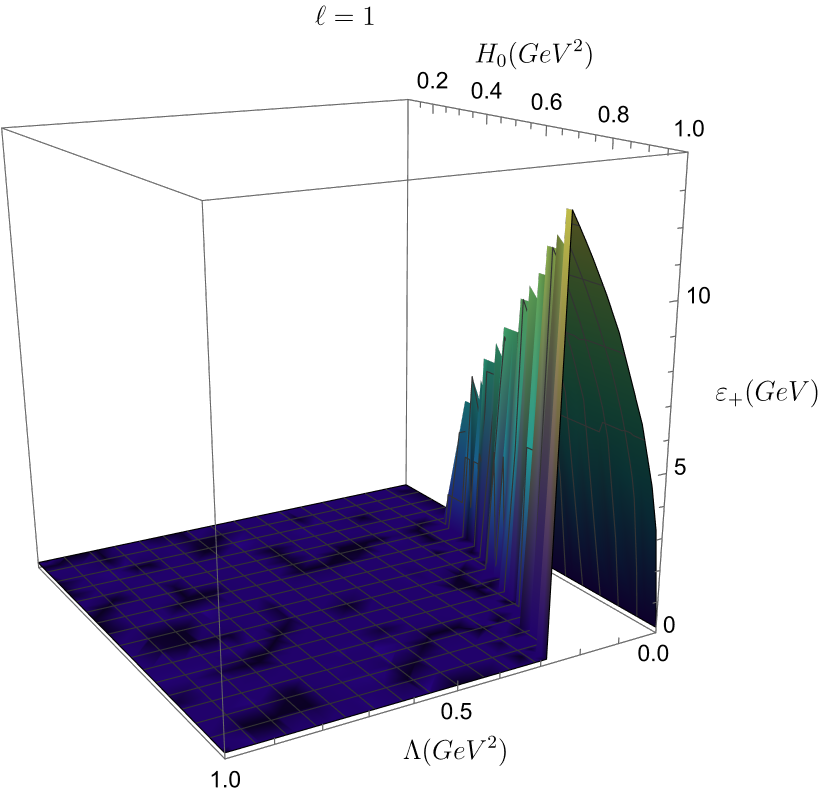}
     \end{subfigure}
     \begin{subfigure}[b]{0.3\textwidth}
         \centering
         \includegraphics[width=\textwidth]{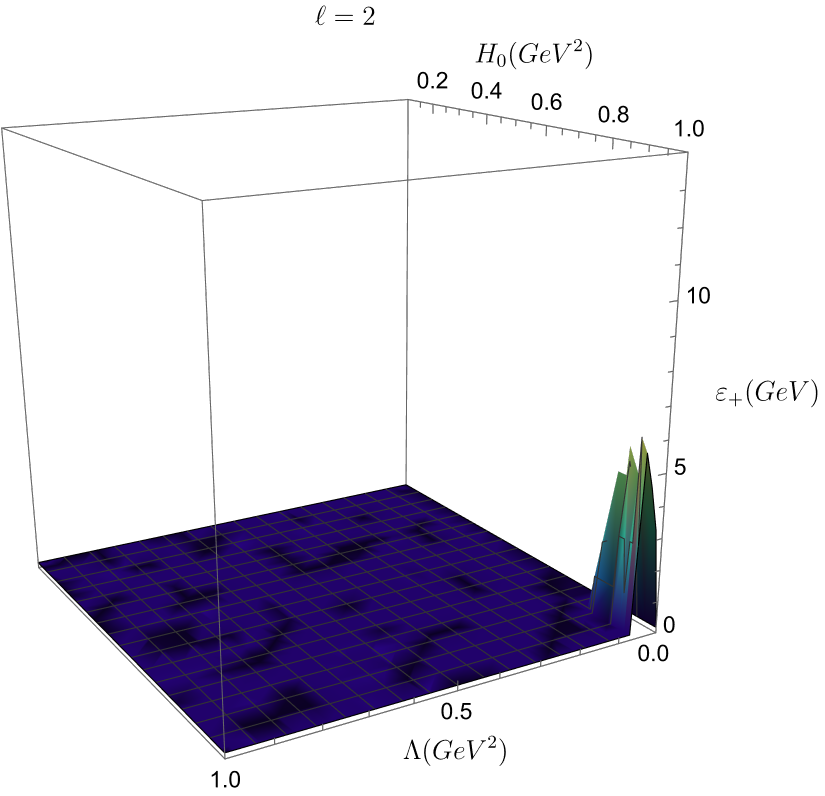}
     \end{subfigure}
    \begin{subfigure}[b]{0.3\textwidth}
         \centering
         \includegraphics[width=\textwidth]{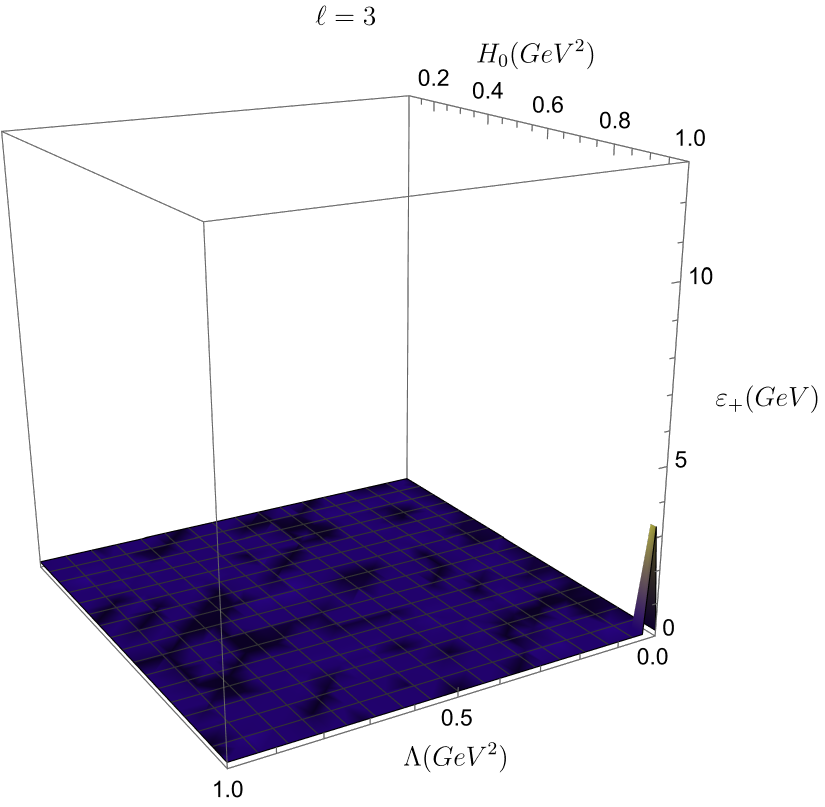}
     \end{subfigure}
        \caption{Graph of the energy spectrum in GeV with respect to the magnetic field intensity $H_{0}$ and the cosmological constant $\Lambda$, for $\ell=1,2,3$, $m=p_z= 0.139$ GeV,  $e=0.3$, $\eta=0.3$ and $n=1000$.}
        \label{f3}
\end{figure}

\begin{figure}[H]
     \centering
     \begin{subfigure}[b]{0.3\textwidth}
         \centering
         \includegraphics[width=\textwidth]{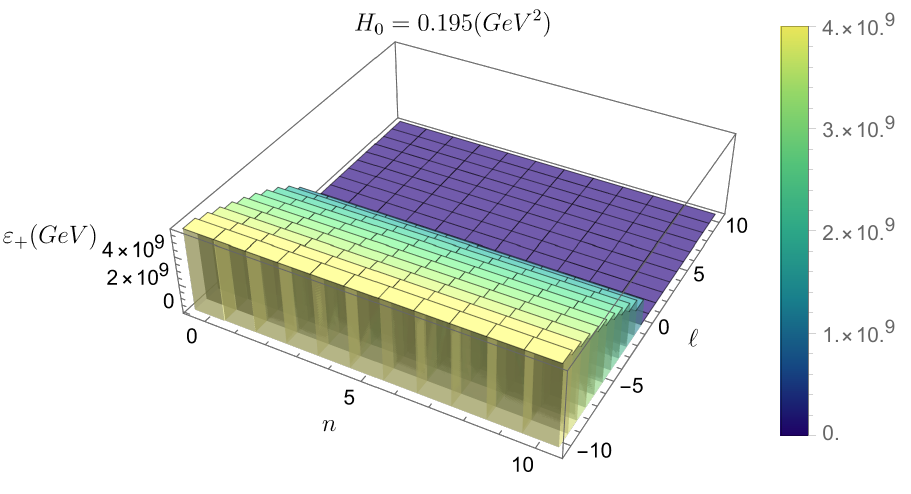}
     \end{subfigure}
     \begin{subfigure}[b]{0.3\textwidth}
         \centering
         \includegraphics[width=\textwidth]{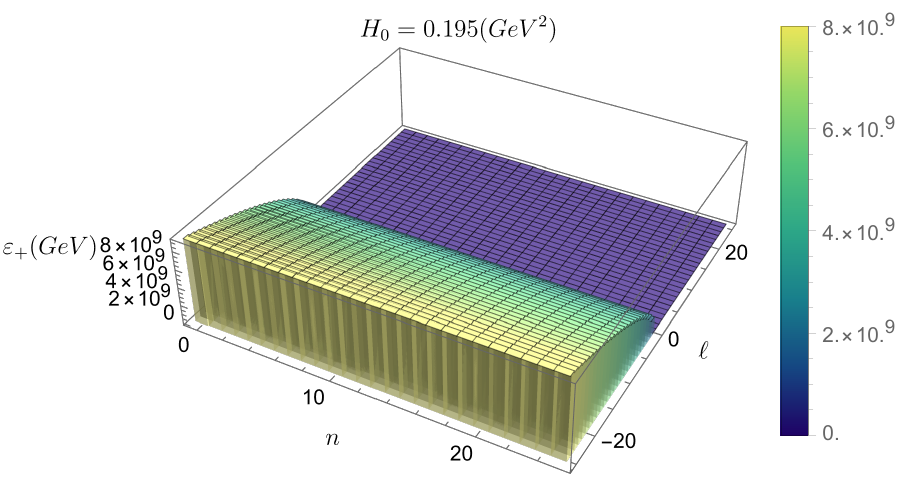}
     \end{subfigure}
    \begin{subfigure}[b]{0.3\textwidth}
         \centering
         \includegraphics[width=\textwidth]{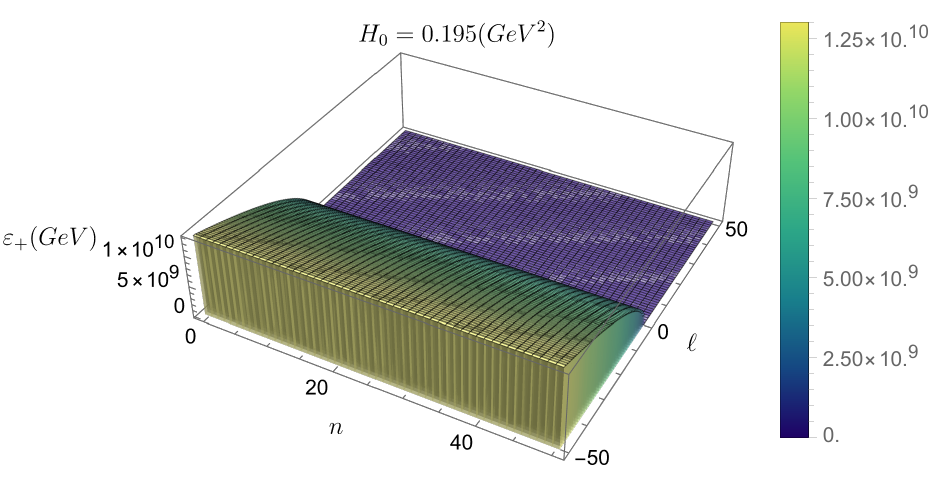}
     \end{subfigure}
        \caption{Plot of the energy spectrum in GeV with respect to the quantum numbers $n$ and $\ell$, with $H_{0}=0.195 \text{ GeV}^{2}$ and the cosmological constant $\Lambda=0.1 \text{ GeV}^{2}$, for $m=p_z= 0.139$ GeV, $e=0.3$, $\eta=0.3$.}
        \label{f4}
\end{figure}

Figures \ref{f3} and \ref{f4} illustrate the energy spectrum behavior using a $\pi^{+}$ meson with a mass of $m = 0.139$ GeV as the test particle \cite{Workman:2022ynf}, \cite{kharzeevmag}. In Fig. \ref{f3} the energy levels are shown in terms of the variation of $\Lambda$ and $H_0$ for $n$= 1000. As we can see, exists a region in the parameters space which the structure of the space time is important and affects the energy levels. In Fig. \ref{f4} the variation of $n$ and $\ell$ is presented and it is possible to see how it can alter the energy levels.

The effect of a scalar Coulomb-like potential on the energy spectrum of scalar bosons was analyzed in this section. Figure \ref{f3} shows the energy levels as a function of $\Lambda$ and $H_0$ for a fixed quantum number $n = 1000$. The energy levels display a complex behavior, with regions where the magnetic field and the cosmological constant have a noticeable influence. Figure \ref{f4} illustrates the energy levels as a function of $n$ and $\ell$ for fixed $\Lambda$ and $H_0$. The energy increases with both $n$ and $\ell$, but the Coulomb-like potential introduces additional shifts in the spectrum. These results show how the Coulomb-like potential modifies the energy levels, particularly near the origin, and how the interplay between the magnetic field, cosmological constant, and the potential affects the quantum states of the particles.

\section{Conclusions}
In this work, we studied spin-0 bosons in a spacetime of a magnetic field with a cosmological constant. The calculations have been made considering the metric proposed in \cite{vzofka2019bonnor} with a modification in the definition of the constants in order to allow an easier interpretation of the limit $\Lambda\rightarrow 0$. We used $\pi^{+}$ as test particles, and in order to have the possibility to achieve more significant effects and strong magnetic fields, of the order of the ones supposed to be produced in high energy heavy ion collisions. The Klein-Gordon equation for this spacetime has been determined and then solved. With these solutions, it was possible to discuss the existence of the Landau levels, find numerical values, and verify their dependence on the considered parameters. In the limit $\Lambda\rightarrow$ 0, we recover the free particle energies, and as $\Lambda$ increases, we get more significant variations. The energy levels also present a behavior similar to the one observed in cosmic string solutions depending on the values of the parameters.
The influence of a scalar Coulomb-type potential has also been verified, and we obtained a solution for small values of $r$ for the purpose of studying quantum particles near the axis of symmetry.   As we can see in Fig. \ref{f1}-\ref{f4}, the effect of the magnetic field $H_0$ may be important in some regions of the graphic. The values of $\Lambda$ and of the other parameters of the theory also influence the results. The calculations presented are determined for a very specific kind of spacetime, but we may imagine that in a region near a source of a strong magnetic field, such as a magnetar, for example, a similar behavior may occur, and this kind of effect may be observed 
for particles near the astrophysical object in accurate experiments.

In \cite{Harding_2006}, \cite{Chatterjee_2021}, it is pointed out that the magnetic field may affect in a significant way the stellar structure. With the results presented in sec. III and IV, we showed that $H_0$ and $\Lambda$ (and the other parameters) are important to determine the values of the Landau levels, and then, this kind of effect will appear in the resulting equation of state, affecting the stellar structure and its physical processes if a stellar model is built with the formulation presented in this work. It is also interesting to remark that the energy-momentum determined in sec. II is not spherically symmetric and depends on $H_0$ and $\Lambda$, if used in the formulation of a star, it will generate anisotropic effects. As we can see, the model proposed in this paper may have a considerable impact on the physics of a star.

This kind of study is very important for the reason that it deals with the influence of the spacetime structure in quantum systems and may provide more information and help to determine the way to correctly formulate a theory in these terms.

\section{Acknowledgements}
We would like to thank CAPES (Process number: 88887.642857/2021-00) and CNPq for the financial support.

\bibliographystyle{ieeetr}
\bibliography{sample}

\begin{thebibliography}{10}

\bibitem{schiff1960possible}
L.~I. Schiff, ``Possible new experimental test of general relativity theory,'' {\em Physical Review Letters}, vol.~4, no.~5, p.~215, 1960.

\bibitem{lense1918einfluss}
J.~Lense and H.~Thirring, ``{\"U}ber den einfluss der eigenrotation der zentralk{\"o}rper auf die bewegung der planeten und monde nach der einsteinschen gravitationstheorie,'' {\em Physikalische Zeitschrift}, vol.~19, p.~156, 1918.

\bibitem{Abbott_2016}
B.~P. Abbot {\em et~al.}, ``Observation of gravitational waves from a binary black hole merger,'' {\em Physical Review Letters}, vol.~116, 2 2016.

\bibitem{einsteingl}
A.~Einstein, ``Lens-like action of a star by the deviation of light in the gravitational field,'' {\em Science}, vol.~84, no.~2188, pp.~506--507, 1936.

\bibitem{SRefsdal_1994}
S.~Refsdal and J.~Surdej, ``Gravitational lenses,'' {\em Reports on Progress in Physics}, vol.~57, p.~117, 2 1994.

\bibitem{VACHASPATI1991258}
T.~Vachaspati, ``Magnetic fields from cosmological phase transitions,'' {\em Physics Letters B}, vol.~265, no.~3, pp.~258--261, 1991.

\bibitem{kharzeevmag}
U.~G\"ursoy, D.~Kharzeev, and K.~Rajagopal, ``Magnetohydrodynamics, charged currents, and directed flow in heavy ion collisions,'' {\em Phys. Rev. C}, vol.~89, p.~054905, 2014.

\bibitem{BZDAK2012171}
A.~Bzdak and V.~Skokov, ``Event-by-event fluctuations of magnetic and electric fields in heavy ion collisions,'' {\em Physics Letters B}, vol.~710, no.~1, pp.~171--174, 2012.

\bibitem{voronyuk}
V.~Voronyuk, V.~D. Toneev, W.~Cassing, E.~L. Bratkovskaya, V.~P. Konchakovski, and S.~A. Voloshin, ``Electromagnetic field evolution in relativistic heavy-ion collisions,'' {\em Phys. Rev. C}, vol.~83, p.~054911, 5 2011.

\bibitem{XU2020135706}
K.~Xu, S.~Shi, H.~Zhang, D.~Hou, J.~Liao, and M.~Huang, ``Extracting the magnitude of magnetic field at freeze-out in heavy-ion collisions,'' {\em Physics Letters B}, vol.~809, p.~135706, 2020.

\bibitem{Skokov20095925}
V.~Skokov, A.~Y. Illarionov, and V.~Toneev, ``Estimate of the magnetic field strength in heavy-ion collisions,'' {\em International Journal of Modern Physics A}, vol.~24, no.~31, p.~5925 – 5932, 2009.
\newblock Cited by: 1010; All Open Access, Green Open Access.

\bibitem{Deng2012}
W.-T. Deng and X.-G. Huang, ``Event-by-event generation of electromagnetic fields in heavy-ion collisions,'' {\em Phys. Rev. C}, vol.~85, p.~044907, Apr 2012.

\bibitem{Chatterjee_2021}
D.~Chatterjee, J.~Novak, and M.~Oertel, ``Structure of ultra-magnetised neutron stars,'' {\em The European Physical Journal A}, vol.~57, Aug. 2021.

\bibitem{dunc1995}
C.~Thompson and R.~C. Duncan, ``{The soft gamma repeaters as very strongly magnetized neutron stars - I. Radiative mechanism for outbursts},'' {\em Monthly Notices of the Royal Astronomical Society}, vol.~275, pp.~255--300, 07 1995.

\bibitem{Kouveliotou:1998ze}
C.~Kouveliotou {\em et~al.}, ``{An X-ray pulsar with a superstrong magnetic field in the soft gamma-ray repeater SGR 1806-20.},'' {\em Nature}, vol.~393, pp.~235--237, 1998.

\bibitem{Harding_2006}
A.~K. Harding and D.~Lai, ``Physics of strongly magnetized neutron stars,'' {\em Reports on Progress in Physics}, vol.~69, p.~2631–2708, Aug. 2006.

\bibitem{chamel2015}
N.~Chamel, Z.~K. Stoyanov, L.~M. Mihailov, Y.~D. Mutafchieva, R.~L. Pavlov, and C.~J. Velchev, ``Role of landau quantization on the neutron-drip transition in magnetar crusts,'' {\em Phys. Rev. C}, vol.~91, p.~065801, Jun 2015.

\bibitem{Chamel_2016}
N.~Chamel, Y.~D. Mutafchieva, Z.~K. Stoyanov, L.~M. Mihailov, and R.~L. Pavlov, ``Landau quantization and neutron emissions by nuclei in the crust of a magnetar,'' {\em Journal of Physics: Conference Series}, vol.~724, p.~012034, jun 2016.

\bibitem{GUTSUNAEV1987215}
T.~Gutsunaev and V.~Manko, ``On the gravitational field of a mass possessing a magnetic dipole moment,'' {\em Physics Letters A}, vol.~123, no.~5, pp.~215--216, 1987.

\bibitem{GUTSUNAEV198885}
T.~Gutsunaev and V.~Manko, ``New static solutions of the einstein-maxwell equations,'' {\em Physics Letters A}, vol.~132, no.~2, pp.~85--87, 1988.

\bibitem{WBBonnor_1954}
W.~B. Bonnor, ``Static magnetic fields in general relativity,'' {\em Proceedings of the Physical Society. Section A}, vol.~67, p.~225, 3 1954.

\bibitem{MELVIN196465}
M.~Melvin, ``Pure magnetic and electric geons,'' {\em Physics Letters}, vol.~8, no.~1, pp.~65--68, 1964.

\bibitem{vzofka2019bonnor}
M.~{\v{Z}}ofka, ``Bonnor-melvin universe with a cosmological constant,'' {\em Physical Review D}, vol.~99, no.~4, p.~044058, 2019.

\bibitem{Parker:1980hlc}
L.~Parker, ``{One-Electron Atom in Curved Space-Time},'' {\em Phys. Rev. Lett.}, vol.~44, no.~23, p.~1559, 1980.

\bibitem{elizalde_1987}
E.~Elizalde, ``Series solutions for the klein-gordon equation in schwarzschild space-time,'' {\em Physical review D: Particles and fields}, vol.~36, pp.~1269--1272, 09 1987.

\bibitem{chandra}
S.~Chandrasekhar, ``The solution of dirac’s equation in kerr geometry,'' {\em Proceedings of the Royal Society of London. A. Mathematical and Physical Sciences}, vol.~349, no.~1659, pp.~571--575, 1976.

\bibitem{Santos:2016omw}
L.~C.~N. Santos and C.~C. Barros~Jr., ``{Scalar bosons under the influence of noninertial effects in the cosmic string spacetime},'' {\em Eur. Phys. J. C}, vol.~77, no.~3, p.~186, 2017.

\bibitem{Santos:2017eef}
L.~C.~N. Santos and C.~C. Barros~Jr., ``{Relativistic quantum motion of spin-0 particles under the influence of noninertial effects in the cosmic string spacetime},'' {\em Eur. Phys. J. C}, vol.~78, no.~1, p.~13, 2018.

\bibitem{Vitoria:2018its}
R.~L.~L. Vit\'oria and K.~Bakke, ``{Rotating effects on the scalar field in the cosmic string spacetime, in the spacetime with space-like dislocation and in the spacetime with a spiral dislocation},'' {\em Eur. Phys. J. C}, vol.~78, no.~3, p.~175, 2018.

\bibitem{Ahmed:2022tca}
F.~Ahmed, ``{Gravitational field effects produced by topologically nontrivial rotating space\textendash{}time under magnetic and quantum flux fields on quantum oscillator},'' {\em Int. J. Mod. Phys. A}, vol.~37, no.~28n29, p.~2250186, 2022.

\bibitem{Ahmed:2023blw}
F.~Ahmed, ``{Klein\textendash{}Gordon oscillator with magnetic and quantum flux fields in non-trivial topological space-time},'' {\em Commun. Theor. Phys.}, vol.~75, no.~2, p.~025202, 2023.

\bibitem{Santos:2019izx}
L.~C.~N. Santos, C.~E. Mota, and C.~C. Barros~Jr., ``{Klein\textendash{}Gordon Oscillator in a Topologically Nontrivial Space-Time},'' {\em Adv. High Energy Phys.}, vol.~2019, p.~2729352, 2019.

\bibitem{Yang:2021zxo}
Y.~Yang, Z.-W. Long, Q.-K. Ran, H.~Chen, Z.-L. Zhao, and C.-Y. Long, ``{The generalized Klein\textendash{}Gordon oscillator with position-dependent mass in a particular G\"odel-type space\textendash{}time},'' {\em Int. J. Mod. Phys. A}, vol.~36, no.~03, p.~2150023, 2021.

\bibitem{Soares:2021uep}
A.~R. Soares, R.~L.~L. Vit\'oria, and H.~Aounallah, ``{On the Klein\textendash{}Gordon oscillator in topologically charged Ellis\textendash{}Bronnikov-type wormhole spacetime},'' {\em Eur. Phys. J. Plus}, vol.~136, no.~9, p.~966, 2021.

\bibitem{Rouabhia:2023tcl}
T.~I. Rouabhia, A.~Boumali, and H.~Hassanabadi, ``{Effect of the Acceleration of the Rindler Spacetime on the Statistical Properties of the Klein\textendash{}Gordon Oscillator in One Dimension},'' {\em Phys. Part. Nucl. Lett.}, vol.~20, no.~2, pp.~112--119, 2023.

\bibitem{Bezerra:2016brx}
V.~B. Bezerra, M.~S. Cunha, L.~F.~F. Freitas, C.~R. Muniz, and M.~O. Tahim, ``{Casimir Effect in the Kerr Spacetime with Quintessence},'' {\em Mod. Phys. Lett. A}, vol.~32, no.~01, p.~1750005, 2016.

\bibitem{Santos:2018jba}
L.~C.~N. Santos and C.~C. Barros~Jr., ``{Rotational effects on the Casimir energy in the space\textendash{}time with one extra compactified dimension},'' {\em Int. J. Mod. Phys. A}, vol.~33, no.~20, p.~1850122, 2018.

\bibitem{Pinho:2023nfw}
E.~O. Pinho and C.~C. Barros~Jr., ``{Spin-0 bosons near rotating stars},'' {\em Eur. Phys. J. C}, vol.~83, no.~8, p.~745, 2023.

\bibitem{Cano:2021qzp}
P.~A. Cano and D.~Pere\~niguez, ``{Quasinormal modes of NUT-charged black branes in the AdS/CFT correspondence},'' {\em Class. Quant. Grav.}, vol.~39, no.~16, p.~165003, 2022.

\bibitem{Sedaghatnia:2019xqb}
P.~Sedaghatnia, H.~Hassanabadi, and F.~Ahmed, ``{Dirac fermions in Som\textendash{}Raychaudhuri space-time with scalar and vector potential and the energy momentum distributions},'' {\em Eur. Phys. J. C}, vol.~79, no.~6, p.~541, 2019.

\bibitem{Guvendi:2022uvz}
A.~Guvendi, S.~Zare, and H.~Hassanabadi, ``{Exact solution for a fermion\textendash{}antifermion system with Cornell type nonminimal coupling in the topological defect-generated spacetime},'' {\em Phys. Dark Univ.}, vol.~38, p.~101133, 2022.

\bibitem{Vitoria:2018mun}
R.~L.~L. Vit\'oria and K.~Bakke, ``{On the interaction of the scalar field with a Coulomb-type potential in a spacetime with a screw dislocation and the Aharonov-Bohm effect for bound states},'' {\em Eur. Phys. J. Plus}, vol.~133, no.~11, p.~490, 2018.

\bibitem{Santos:2015esa}
L.~C.~N. Santos and C.~C. Barros~Jr., ``{Dirac equation and the Melvin Metric},'' {\em Eur. Phys. J. C}, vol.~76, no.~10, p.~560, 2016.

\bibitem{Workman:2022ynf}
R.~L. Workman {\em et~al.}, ``{Review of Particle Physics},'' {\em PTEP}, vol.~2022, p.~083C01, 2022.

\bibitem{dosch1971kleins}
H.~G. Dosch, V.~Muller, and J.~Jensen, ``Kleins paradox,'' {\em Physica Norvegica}, vol.~5, pp.~3--4, 1971.

\bibitem{bergerhoff1994scalar}
B.~Bergerhoff and G.~Soff, ``Scalar potentials and the dirac equation,'' {\em Zeitschrift f{\"u}r Naturforschung A}, vol.~49, no.~11, pp.~997--1012, 1994.

\bibitem{figueiredo2012relativistic}
E.~Figueiredo~Medeiros and E.~Bezerra~de Mello, ``Relativistic quantum dynamics of a charged particle in cosmic string spacetime in the presence of magnetic field and scalar potential,'' {\em The European Physical Journal C}, vol.~72, no.~6, p.~2051, 2012.

\bibitem{ronveaux1995heun}
A.~Ronveaux, ``Heun's differential equations,'' {\em (No Title)}, 1995.

\bibitem{Cunha:2016uch}
M.~S. Cunha, C.~R. Muniz, H.~R. Christiansen, and V.~B. Bezerra, ``{Relativistic Landau Levels in the Rotating Cosmic String Spacetime},'' {\em Eur. Phys. J. C}, vol.~76, no.~9, p.~512, 2016.

\bibitem{vieira2015quantum}
H.~S. Vieira and V.~B. Bezerra, ``Quantum newtonian cosmology and the biconfluent heun functions,'' {\em Journal of Mathematical Physics}, vol.~56, no.~9, 2015.

\end{thebibliography}

\end{document}